\documentclass[aps,prb]{revtex4}   
\include{amsmath}
\input{epsf}
\tighten
\begin{document}
\title{{\bf Protein folding in a force-clamp}}
\author{{\bf Marek Cieplak$^1$ and P. Szymczak $^2$}}

\address{
$^1$Institute of Physics, Polish Academy of Sciences,
Al. Lotnik\'ow 32/46, 02-668 Warsaw, Poland\\
$^2$Institute of Theoretical Physics, Warsaw University,
ul. Ho\.za 69, 00-681 Warsaw, Poland}



\begin{abstract}
{Kinetics of folding of a protein held in a force-clamp are compared
to an unconstrained folding. The comparison
is made within a simple topology-based dynamical model of ubiquitin.
We demonstrate that the experimentally observed variations in the
end-to-end distance reflect microscopic events during folding.
However, the folding scenarios in and out of the force-clamp are distinct.}
\end{abstract}

\maketitle

Recent advances in nano-technology have enabled manipulation of single
biomolecules, especially by means of the atomic force microscopy (AFM). The
manipulation usually involves mechanical stretching and monitoring of the
force of resistance as a function of displacement of the AFM tip. In 2001,
Oberhauser et al. \cite{clampober} have developed a force-clamp --
a variant of AFM with an electronic adjustment
of the tip displacement so that a constant
pulling force is maintained. This technique allows one to measure the force
dependence of the unfolding probability and has been used to probe the
mechanical stability of two different domains of titin.
The force-clamp microscopy has been subsequently developed by Fernandez
and Li \cite{FernandezLi} to monitor the folding trajectory of a single
protein that is first stretched by a constant unfolding force and
then suddenly submitted to a substantially reduced force. The first tests
on polyubiquitin have demonstrated a structured time dependence of the
end-to-end distance, $L$. What this behavior corresponds to
microscopically remains to be elucidated.\\

In this paper we ask what one can learn from monitoring $L$ during
folding of a protein under a small stress and, in
particular, is this process related
to folding that is taking place without any mechanical constraints?
We address these questions theoretically by performing molecular dynamics
simulations in a coarse-grained topology-based
model \cite{Goabe}. Such a model is
ideally suited to study conceptual questions about large conformational
changes because it makes relevant time scales accessible to computations.
\\

We focus on ubiquitin and two-ubiquitin since this case relates
to the experimental studies \cite{FernandezLi,Schlierf}.
A single ubiquitin consists of 76 amino acids
and its structure is deposited in the Protein Data Bank \cite{PDB}
with a code 1ubq.
The two-ubiquitin is modeled
by linking two ubiquitins in a series through an extra peptide bond.
We follow the implementation of the model along the lines
outlined in refs. \cite{Hoang}. \\

In short, a protein is represented by
a chain of C$^{\alpha}$ atoms that are
tethered by harmonic potentials with minima at 3.8 {\AA}.
The effective self-interactions between the atoms are
either purely repulsive or are minimum-endowed-contacts of the
Lennard-Jones type,
$V_{ij} =
4\epsilon \left[ \left( \frac{\sigma_{ij}}{r_{ij}}
\right)^{12}-\left(\frac{\sigma_{ij}}{r_{ij}}\right)^6\right]$.
The length parameters $\sigma _{ij}$ are chosen so that the potential
minima correspond, pair-by-pair, to the
experimentally established native distances between
the C$^{\alpha}$ atoms in amino acids in the pair.
The distinction between the two kinds of the interactions is
established based on the absence or presence of overlaps
between all other atoms in the $i$ and $j$ amino acids in the native
conformation. The effective geometry of atoms in the tests for
overlaps is assigned following a procedure advanced
by Tsai et al. \cite{Tsai}.
The repulsive interactions are described by the $r_{ij}^{-12}$ part
of the Lennard-Jones potential combined with a constant shift term
that makes the potential vanish smoothly at $\sigma =5$ {\AA}.
It should be noted that the specificity of a protein is
contained in the length parameters $\sigma _{ij}$ and not in the
energy parameter, $\epsilon$. The energy parameter is taken
to be uniform and its effective value for titin and ubiquitin
appears to be of order 900 K so the reduced temperature,
$\tilde{T}=k_BT/\epsilon \;$ of 0.3
($k_B$ is the Boltzmann constant and $T$ is temperature)
should be close to the room temperature
value \cite{Pastore}. All of the simulations
reported here were performed at this temperature.
In our stretching simulations, the N-terminus of the protein is
attached to harmonic springs of elastic
constant $k$=0.06 $\epsilon /${\AA}$^2$.
The C-terminus is pulled by a constant force, $F$. The
dimensionless force, $F\sigma/\epsilon$, will be denoted by $\tilde{F}$.\\

Thermostating and mimicking some other effects of the solvent
is provided by the Langevin noise. An equation of motion
for each C$^{\alpha}$ reads then
$m\ddot{{\bf r}} = -\gamma \dot{{\bf r}} + F_c + \Gamma $, where
$F_c$ is the net force on an atom due to the molecular potentials
and $\Gamma$ is a Gaussian noise term with dispersion
$\sqrt{2\gamma k_B T}$.
The damping constant $\gamma$ is taken to be equal to $2m/\tau$
and the dispersion of the random forces is equal to
$\sqrt{2\gamma k_B T}$.
This choice of $\gamma$ corresponds to a situation in which the
inertial effects are small\cite{Hoang}
but the damping action is not yet as strong as in water.
The more realistic damping is stronger by a factor of order 25
which extends the effective time scales by the same factor
since the dependence of the folding times on $\gamma$ is linear
\cite{Hoang}.
The equations
of motion are solved by a fifth order predictor-corrector scheme.
The molecular dynamics time evolution is governed by the time unit
$\tau =\sqrt{m \sigma^2 / \epsilon} \approx 3$ps where
$m$ is the average mass of the amino acids. For our model of ubiquitin,
the reduced temperature of melting is 0.23 and that of optimal folding 0.28
which indicates good unfrustrated kinetics and
two-state folding behavior \cite{Hoang}.
In an unfolded state there are no contacts and folding means increasing
the number of contacts that get established. A contact between
amino acids $i$ and $j$ is said to be established
if the corresponding distance $r_{ij}$ becomes less
than $1.5 \sigma _{ij}$ (close to the inflection point of the
Lennard-Jones potential). 
The folding time $t_{fold}$ is defined as the first time at which all contacts are established simultaneously. 
(The condition of simultaneity necessarily
involves longer time scales than those connected to single contact events).\\

Figure 1 illustrates examples of the experimental-like protocol that is
adopted when studying force-clamped folding:
a strong force of $\tilde{F}$=2 is applied to induce unfolding and once
this is accomplished, the force is reduced to a smaller value of
$\tilde{F}'$=0.3 to generate refolding; after refolding, the second
such cycle of stretching and refolding is initiated, etc.
Force induced unfolding -- the first leg of the protocol --
has been analyzed for ubiquitin theoretically before \cite{Makarov,Szymczak}
and now we focus on what follows after the force
quench.
For single ubiquitin (the top two panels), there is a single jump in
$L$ during stretching, whereas for two-ubiquitin (the bottom panel)
$L$ increases in two steps, indicating a serial nature of unwinding --
domain after domain. Figure 2 shows that the bigger the $\tilde{F}'$,
the longer the mean refolding time, $<t_{fold}>$. This also means
the longer lasting intervals between noticeable shortenings of $L$.
As it is seen in Fig. 1, $L$ shrinks in an almost
continuous fashion for $\tilde{F}'$=0.3  whereas
the steps are more pronounced and long lasting for $\tilde{F}'$=0.36
because the larger stretching force generates a stronger impediment to folding.\\

It should be noted that even when $\tilde{F}'$=0, the folding process
is not identical to a clamp-free folding because the N-terminus
is connected to a spring that is anchored. The inset of Figure 2
shows that fixing one end of a protein delays folding nearly by a factor of 2.
This seems to imply that diffusion-limited processes play an important
role in the ubiquitin folding.
Namely, according to the classical Smoluchowski result \cite{smol}
the binding rate for a diffusion limited process is $k_s = 4 \pi Da$, where
$D$ is the relative diffusion coefficient of a
pair of reactants (amino acids) which are to make a contact once within
a final distance of $a$.
If one of them is immobile, $D$ is halved and the time needed to
form a contact increases accordingly.
This argument is also consistent with the diffusion-collision
model of Karplus and Weaver \cite{Weaver}.
It is worth noting that for a shorter system, such as a single
helix \cite{helix}, the situation is reversed: helix with a fixed end folds
slightly faster than a free one. However, the diffusion is not a limiting
process in the helix, since the sequential distance of aminoacids in the native
contact in the helix does not exceed 4 and the process is energy driven.
The factor of two difference in the folding time between free and clamped
ubiquitin is also in a good agreement with the data obtained by
Fernandez and Li \cite{FernandezLi} when an extrapolation to
$F'=0$ is done and compared to the free ubiquitin folding
times (see eg. \cite{KrantzSos} and references therein).
Sosnick \cite{Sosnick} interprets this difference in folding
times as having origin in random nature of aggregation
that partially unfolded molecules of ubiquitin
may participate in when in a dense solution.
The intimacy of protein chains would then give rise to the
additional resistance to folding and hence to the longer times.
Our results suggest that the aggregation mechanism
is not necessary: single molecules themselves  give rise to the
phenomenon.\\

The distributions of the folding times  both for free and clamped end are
well fitted by the log-normal distribution except for large times where a
transition to power-law-like tail is seen.
Similar distributions were reported by Zhou et al. \cite{Zhou2} for the
free folding of $\beta$-hairpin fragment of protein G whereas the power
law tails were predicted theoretically based on the energy
landscape theory \cite{Lee} and the hierarchically constrained dynamics
model \cite{Skoro}.

It should be noted that the shape of the distribution reflects
variations in the time scale of folding which are due to variations
in the unfolded initial states,
random character of the Langevin noise,
and some variety in the shapes of final conformations
that result when  folding is declared accomplished.
Such distributions exist for many simple proteins no matter which model
of a protein is adopted. \\

There is a related implementation of the Go-like model \cite{Veitshans,Clementi} in which instead of the
Lennard-Jones contact potentials
combined with the chirality terms one considers the 10-12 contact
potentials combined with potentials that involve the bond and
dihedral angles. It has been shown \cite{Clementi} that the other
implementation is endowed with an explicit two-state
behavior since its free energy, at least for some proteins,
has a two-minima form when plotted against the fraction, $Q$, of
the established native contacts. In particular Zhang et al. \cite{Zhang}
have demonstrated the
thermodynamic two-state behavior for ubiquitin and
Li et al. \cite{Liklimov} have used it to study refolding upon
force quench of titin and to show dependence on initial conditions.
The technical differences between the two implementations do not
translate into any qualitative differences in kinetics and
thermodynamics \cite{angular} and our choice is motivated
by a better computational efficiency.\\

Force-clamping results not only in longer folding but also in channeling
the process through a modified pathway even if $\tilde{F}'$=0. The
kinetics of folding can be quantified with the use of the so called
scenario diagrams \cite{Hoang,coop} in which one
plots an average time to establish a contact, $t_c$, against the contact
order, i.e. against the sequential distance, $j-i$, between the
amino acids that form a native contact.
The left panels of Figure 3 compare the folding scenarios for the unconstrained
and anchored (at the N terminus) single ubiquitin.
The free case reveals folding that is fairly monotonic as a function of
the contact order: the short-range contacts are established the first
and the long-range contacts the last. When one end is held fixed, the
sequence of events splits into more identifiable branches
so that contacts of a given contact order are established at up to
three distinct time scales. In addition, the order of events gets
overturned. For instance, the hairpin at the N-terminus
(solid squares) gets
established now near the completion of folding,
instead of at the beginning,
despite the small sequential distances in the hairpin.
Another difference is that there is no longer any time
separation between linking the segment (17-27) with (51-59) (solid triangles) and
connecting (36-44) to the strand (65-72) (open circles). The panels on the
right hand side of Figure 3 make a similar comparison for a two-ubiquitin
tandem arrangement and demonstrate that in the clamp-free folding
process the folding events involve the two ubiquitins simultaneously.
On the other hand, if one end is clamped, then the free-end domain
folds first, as it can diffuse faster, and the domain that is close to the clamp
folds next. Thus fixing one end induces seriality in folding. (We find that
the same holds when a non-zero $\tilde{F}'$ is applied).\\

The left panel of Figure 4 shows
that application of the force at the C-terminus
affects folding of a single ubiquitin even  further.
It restores the time separation
between the events corresponding to the solid triangles and open
circles; it makes the contacts with the helix (open triangles)
take place at the very beginning of folding, instead of
in the middle; it keeps establishing the hairpin throughout the process
until the very end. We conclude that folding in a force-clamp
is distinct both from folding that is not restricted mechanically
and from folding that is partially restricted.\\

Another important aspect of the force-clamped folding is that its
time scales are extended -- the bigger the $\tilde{F}'$, the longer the
folding time and, in particular, larger time intervals between
individual events.
This makes it easier for the atomic force microscopy
to sense major events in folding, especially those that result
in large time gaps between a next stage of contact formation.
The right-hand panel of Figure 4 shows the distance $L$ as a function
of time for  $\tilde{F}'$.
The way the panel is plotted is rotated by 90$^o$ so that the time
axis is parallel to that on the left-hand panel and it spans
the same duration.
It is seen that $L$ varies in a
discontinuous fashion forming a pattern of "punctuated equilibria".
Furthermore, the jumps are very well correlated with the
scenario diagram: the time gaps between establishment
of subsequent contacts are reflected in a nearly constant
value of $L$ and a rapid chain of events make the $L$
suddenly shorter.
When the number of ubiquitins is larger than one (see Figure 1),
the rapid changes in $L$ are less pronounced since the unfolded modules
act as soft entropic springs \cite{Hummer} whose length fluctuations
tend to mask the decrease in the total $L$. As noted by Best and
Hummer \cite{Hummer}, resolving kinetic events could be enhanced
by operating at higher forces that reduce fluctuations. We concur
and also point to the beneficial effects of the increased time scales
in resolving the folding events.\\

In conclusion, results on our dynamical model suggest that
monitoring the end-to-end distance in a force-clamp microscope
does probe folding in a meaningful way. However,
the folding process itself is different from that taking place
without any mechanical constraints.\\

M.C. appreciates discussions with Julio Fernandez and Andrzej Sienkiewicz.
This work was funded
by the Ministry of Science in Poland (grant 2P03B 03225).


\newpage

\centerline{FIGURE CAPTIONS}

\begin{description}

\item[Fig. 1. ]
Examples of $L$ vs. time trajectories in two cycles of
the force $\tilde{F}$ varying
in the following fashion: 2--$\tilde{F}'$--2--$\tilde{F}'$, where
the values of $\tilde{F}'$ are indicated.
The top two panels are for ubiquitin and the bottom panel
for two-ubiquitin. The dashed line in the upper panel shows the time dependence of the streching force in this case. The scale for $\tilde{F}$ coincides with that of $L/100$.

\item[Fig. 2. ]
The mean folding time, $<t_{fold}>$, as a function of $\tilde{F}'$.
 The inset shows
distributions of folding times for 1ubq with one terminus fixed
(the shaded histogram) and with free ends.
The fits are to a log-normal distribution
$\frac{1}{\sqrt{2\pi}\sigma (t-t_0)}
\exp{(-\frac{ln^2(\frac{t-t_0}{m})}{2\sigma ^2})}$.
The values of $t_0/\tau$, $\sigma$, and $m/\tau$ are 1050, 0.37, 820
and 480, 0.9, 215 respectively. The refolding time is measured from
the drop in the force.

\item[Fig. 3. ]
Folding scenarios for a single ubiquitin (the left hand panel)
and for two ubiquitins connected in tandem (the right hand panels) as
averaged over 100 trajectories.
The top panels correspond to unconstrained processes
whereas the bottom panels to processes in which the N-terminus is fixed.
The symbols assigned to specific contacts are the same in the panels on
the left. Open circles, open triangles, open squares, open pentagons,
solid triangles, and solid
solid squares correspond to contacts (36-44)--(65-72), (12-17)--(23-34),
[(1-7),(12-17)]--(65,72), (41-49)--(41-49), (17-27)--(51-59), (1-7)--(12-17)
respectively. The crosses denote all other contacts.
The segment (23-34) corresponds to a helix. The two $\beta$-strand
(1-7) and ((12-17) form a hairpin. The remaining $\beta$-strands
are (17-27), (41-49), and (51-59). In the panels on the right,
we assign open circles to all contacts that exist in ubiquitin on the
N-terminal side of the tandem arrangement whereas the solid triangles
to contacts in the other ubiquitin.

\item[Fig. 4.]
The left panel shows the scenario of folding for $\tilde{F}'$=0.36.
The symbols used are like in the left panels of Figure 3.
In a tandem arrangement of two ubiquitin at this force, there is a serial
folding but the choice of the first ubiquitin to fold is random (not shown).
The right panel shows the
corresponding values of $L$, together with the snapshots of the
conformations. The top snapshot is for the native conformation.
The constant force is applied to the terminus shown on the right-hand
side of the snapshot. The left-hand side terminus is attached to
a fixed spring.

\end{description}

\begin{figure}
\epsfxsize=7in
\centerline{\epsffile{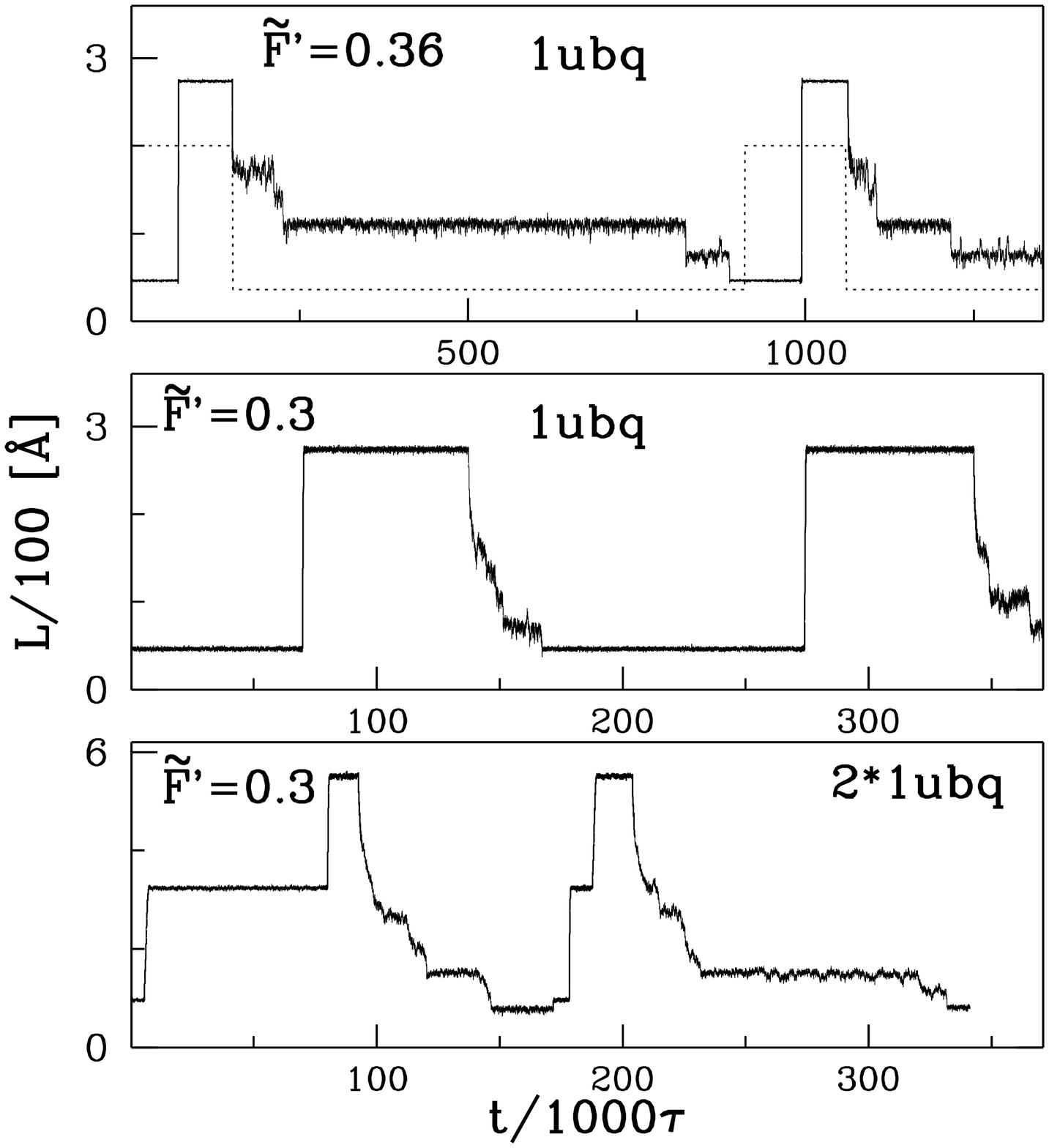}}
\vspace*{3cm}
\caption{ }
\end{figure}

\begin{figure}
\epsfxsize=7in
\centerline{\epsffile{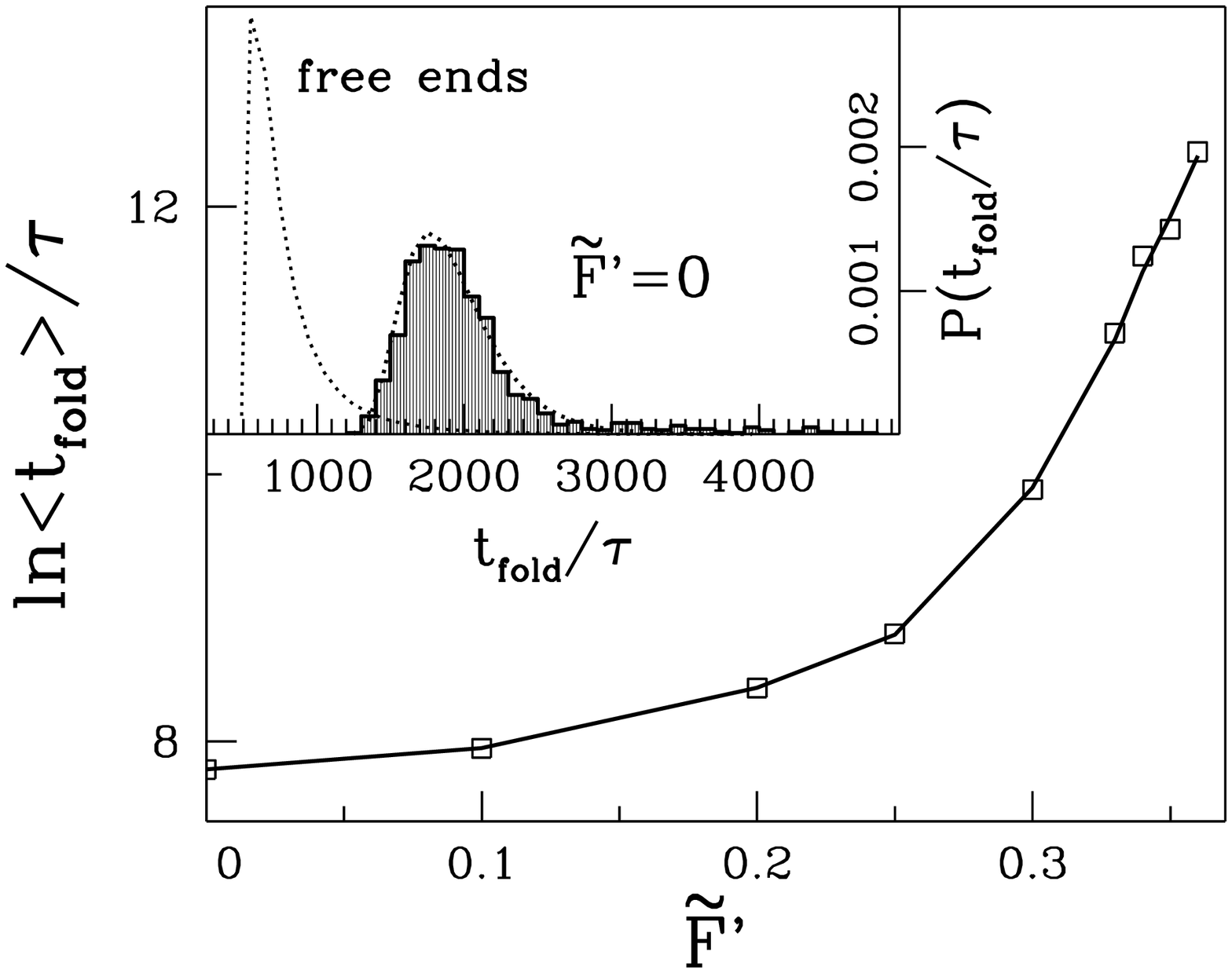}}
\vspace*{3cm}
\caption{ }
\end{figure}

\begin{figure}
\epsfxsize=7in
\centerline{\epsffile{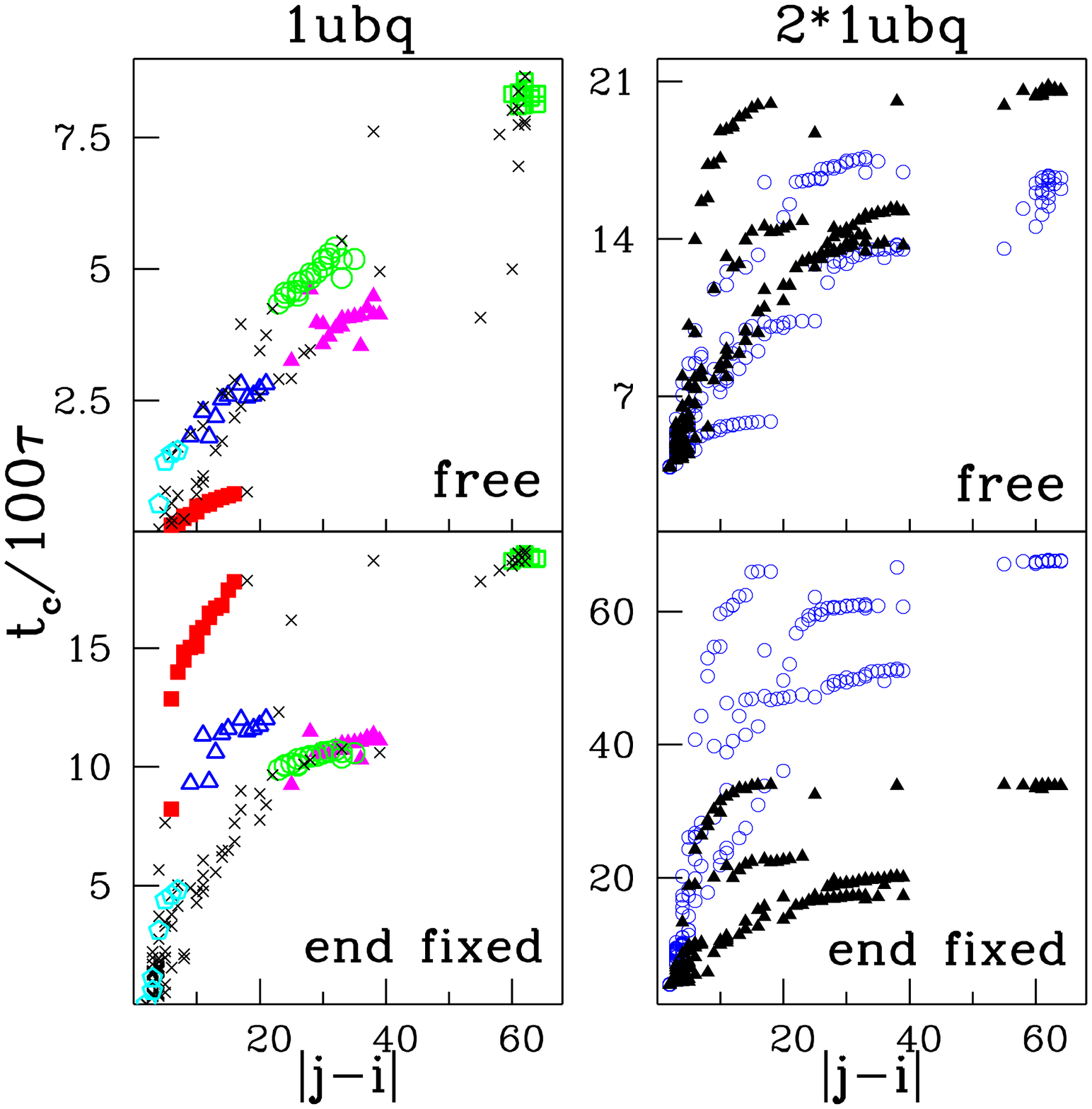}}
\vspace*{3cm}
\caption{ }
\end{figure}

\begin{figure}
\epsfxsize=6.6in
\centerline{\epsffile{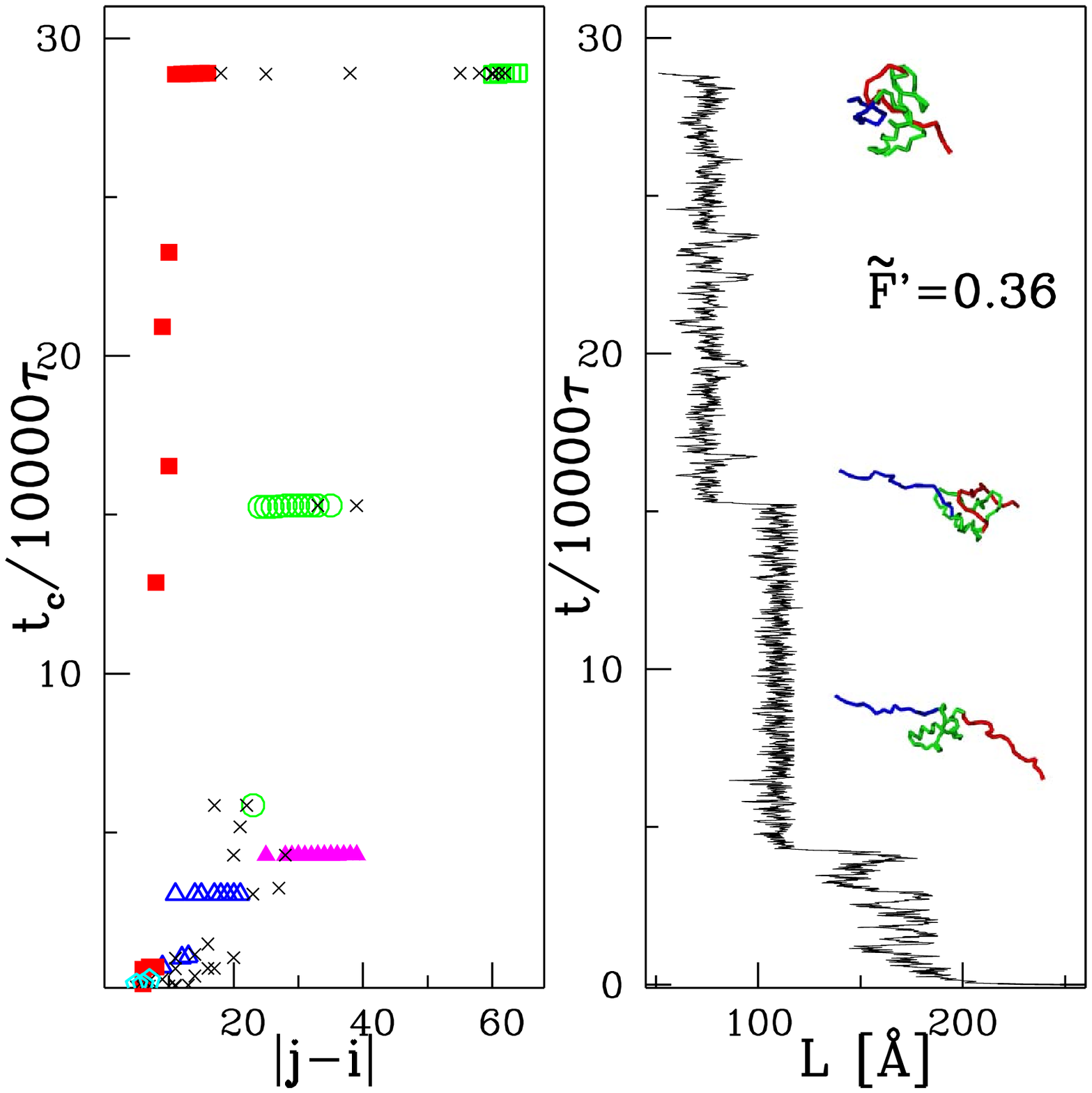}}
\caption{ }
\end{figure}


\begin{thebibliography}{99}

\bibitem{clampober}
A. F. Oberhauser, P. K. Hansma, M. Carrion-Vazquez, and J. M. Fernandez,
{\it Proc. Natl. Acad. Sci. USA} {\bf 98}, 468 (2001).

\bibitem{FernandezLi}
J. M. Fernandez and H. Li,
{\it Science} {\bf 303} 1674 (2004).

\bibitem{Schlierf}
M. Schlierf, H. Li, J. M. Fernandez
{\it Proc. Natl. Acad. Sci. USA}, {\bf 101}, 7299 (2004)

\bibitem{Goabe}
H. Abe and N. Go,
{\it Biopolymers} {\bf 20} 1013 (1981);
S. Takada,
{\it Proc. Natl. Acad. Sci. (USA)} {\bf 96} 11698 (1999).


\bibitem{PDB}
H.M. Berman, J. Westbrook, Z. Feng, G. Gilliland, T.N. Bhat, H. Weissig,
I.N. Shindyalov, P.E. Bourne:
{\it Nucleic Acids Research}, {\bf 28} 235 (2000).   

\bibitem{Hoang}
T. X. Hoang and M. Cieplak,
{\it J. Chem. Phys.} {\bf 113}, 8319-8328 (2000);
M. Cieplak and T. X. Hoang,
{\it Biophysical J.} {\bf 84} 475 (2003);
M. Cieplak, T. X. Hoang and M. O. Robbins,
{\it Proteins: Struct. Funct. Bio.} {\bf 56} 285 (2004).


\bibitem{Tsai}
J. Tsai, R. Taylor,  C. Chothia, and M. Gerstein,
{\it J. Mol. Biol.} {\bf 290} 253 (1999).

\bibitem{Pastore}
M. Cieplak, A. Pastore and T. X. Hoang,
{\it J. Chem. Phys.} {\bf 122} 054906 (2004);
M. Cieplak and P. E. Marszalek,
{\it J. Chem. Phys.} {\bf 123}, 194903 (2005).

\bibitem{evans}
E. Evans
{\it Faraday Discuss.} {\bf 111}, 16 (2003)

\bibitem{Makarov}
P. C. Li, and D. E. Makarov,
{\it J. Chem. Phys.} {\bf 121}, 4826 (2004).

\bibitem{Szymczak}
P. Szymczak and M. Cieplak,
{\it J. Phys.: Cond. Mat.} {\bf 18} L21 (2006)

\bibitem{smol}
M. Smoluchowski,
{\it Z. Phys. Chem.} {\bf 92}, 129 (1917);     
H.-X. Zhou
{\it J. Mol. Recognit.} {\bf 17}, 368 (2004).

\bibitem{Weaver}
M. Karplus and D. L. Weaver,
{\it Nature} {\bf 260}, 404 (1976).   

\bibitem{helix}
T. X. Hoang and M. Cieplak,
{\it J. Chem. Phys.} {\bf 112}, 6851 (2000).    

\bibitem{KrantzSos}
B. A. Krantz, T. R. Sosnick
{\it Biochemistry} {\bf 39}, 11696 (2000).          

\bibitem{Sosnick}
T. R. Sosnick,
{\it Science} {\bf 306}, 411 (2004)

\bibitem{Zhou2}
Y. Zhou, C. Zhang, G. Stell, J. Wang,
{\it J. Am. Chem. Soc.} {\bf 125}, 6300 (2003).

\bibitem{Lee}
C.-L. Lee, C.-T. Lin, G. Stell, J. Wang,
energy
{\it Phys. Rev. E} {\bf 67}, 041905 (2003).

\bibitem{Skoro}
M. Skorobogatiy, H. Guo, M. Zuckermann
{\it J. Chem. Phys.} {\bf 109}, 2528 (1998).

\bibitem{Veitshans}
T. Veitshans, D. Klimov, and D. Thirumalai,
{\it Folding Des.} {\bf 2}, 1-22 (1997).

\bibitem{Clementi}
C. Clementi, H. Nymeyer, and J. N. Onuchic.
{\it J. Mol. Biol.} {\bf 298}, 937 (2000).

\bibitem{Zhang}
J. Zhang, M. Qin and W. Wang,
{\it Proteins: Struct. Funct. Bio.} {\bf 59}, 565 (2005).

\bibitem{Liklimov}
M. S. Li, C-K. Hu, D. Klimov, and D. Thirumalai,
{\it Proc. Natl. Acad. Sci. (USA)} {\bf 93} (2006).

\bibitem{angular}
M. Cieplak and Trinh Xuan Hoang,
{\it Physica A} {\bf 330}, 195 (2003).

\bibitem{coop}
M. Cieplak,
{\it Phys. Rev. E} {\bf 69}, 031907 (2004).

\bibitem{Hummer}
R. B. Best and G. Hummer,
{\it Science} {\bf 308}, 498B (2005).

\end{thebibliography}
\end{document}